\documentclass[11pt,a4paper]{article}
\usepackage[dvipdfmx]{graphicx,xcolor}
\usepackage[fleqn]{amsmath}
\usepackage{amsthm}
\usepackage{newtxtext}
\usepackage[varg]{newtxmath}
\usepackage{latexsym}
\usepackage{algorithm}
\usepackage{algorithmic}

\usepackage[mathlines]{lineno}
\usepackage{braket}
\usepackage{graphicx}
\usepackage{latexsym}
\usepackage{autobreak}
\usepackage{nccmath}
\usepackage{bmpsize}
\usepackage{here}
\title{Advance sharing for stabilizer-based quantum secret sharing schemes}
\author{Mamoru Shibata\thanks{Department of Computer Science, National Institute of Technology, m\_shibata@tokyo-ct.ac.jp}}

\theoremstyle{definition}
\newtheorem{example}{Example}
\newtheorem{lemma}{Lemma}

\newtheorem{remark}{\textbf{Remark}}

\begin{document}
\maketitle
\begin{abstract}
    In stabilizer-based quantum secret sharing schemes, it is known that some shares can be distributed to participants before a secret is given to the dealer. 
    This distribution is known as advance sharing.
    It is already known that a set of shares is advance shareable only if it is a forbidden set.
    However, it was not known whether any forbidden set is advance shareable. 
    We provide an example of a set of shares such that it is a forbidden set but is not advance shareable in the previous scheme.
    Furthermore, we propose a quantum secret sharing scheme for quantum secrets such that any forbidden set is advance shareable.
\end{abstract}

\section{Introduction}
To protect important information from destruction or loss, we should not store it in one place, but we should store copies of it across multiple places and media. 
However, if the important information is secret, this strategy clearly increases the risk of information leakage.
A revolutionary method to solve this problem is the secret sharing (SS), which was invented independently by Shamir \cite{SS_Shamir_1979} and Blakley \cite{SS_Blakley_1979} in 1979.
SS is a cryptographic scheme to encode a secret to multiple shares being distributed to participants, 
so that certain sufficiently large sets of participants can reconstruct the secret from their shares.
A set of participants that can reconstruct the secret is called a qualified set, and 
a set of participants that can gain no information about the secret is called a forbidden set.
The set of qualified sets and that of forbidden sets are called an access structure \cite{stinson2005cryptography}. 
In quantum information theory, Hillery et al.\ \cite{QSS_Hillery_1999} and Cleve et al.\ \cite{QSS_Cleve_1999} simultaneously presented the quantum secret sharing (QSS) scheme in 1999.
Cleve et al.\ clarified the relationships between QSS and quantum error-correcting codes. 
In that relations, a share of QSS is each qudit of a codeword in a quantum error-correcting code \cite{QSS_Cleve_1999}.
The well-known classes of quantum error-correcting codes are the CSS codes \cite{Calderbank_1996,Steane_1996}, 
the stabilizer codes \cite{gottesman1997stabilizer} that include the CSS codes as a special case.
QSS constructed from a stabilizer code had been already studied \cite{Marin_StabilizerQSS,Markham_StabilizerQSS,Sarvepalli_StabilizerQSS}.
Stabilizer-based QSS is important because it can realize access structures that cannot be realized by quantum SS based on CSS codes.
For example, only the $[[5,1,3]]$ binary stabilizer codes can realize QSS distributing 1 qubit of secret to 5 participants receving 1-qubit shares and allowing only 3 or more participants to reconstruct the secret.

In traditional secret sharing schemes, 
a dealer is assumed to be always able to communicate with all participants. 
However, it is sometimes difficult for the dealer to distribute shares 
when the dealer obtains a secret.
For example, a situation where some participants will be in locations where communication is not possible.
To solve this problem, the dealer distribute shares to these participants while the dealer can communicate with them. 
To realize this distribution, the dealer needs to be capable of distributing shares to some participants before a given secret.
We call a distribution of shares to some participants before a given secret ``advance sharing'' and a set of shares 
that can be distributed in advance is called ``advance shareable'' \cite{Miyajima_2022}. 

In QSS, advance sharing schemes utilizing enatanglement-assisted quantum error-correcting codes (EAQECCs) \cite{AdvanceSharing_Shibata,masumori2024advance} and Quantum Masker \cite{Lie_2020} have been known.
The advance sharing scheme utilizing EAQECCs works as follows:
\begin{enumerate}
    \item A dealer prepares some pairs of maximally entangled states and distributes halves of these pairs to participants in adnvance shareable set.
    \item The dealer encodes a quantum secret into a codeword of an EAQECC. 
    \item The dealer distributes each qudit of the encoded state to the remaining participants.
\end{enumerate}
It is known that a set of shares is an advance shareable set only if it is a forbidden set. 
In QSS, it is desirable for the advance shareable set to be large. 
However, it was not known whether any forbidden set is advance shareable. 
Therefore, if participants for advance sharing are determined before deciding the access structure, it is not known what access structures could be constructed.
Hence, when constructing QSS, it was necessary to repeatedly pick an access structure and then check whether the advance shareable set for that access structure was appropriate.

In this paper, we provide an example such that a set of shares is a forbidden set but not an advance shareable set by EAQECCs \cite{AdvanceSharing_Shibata}.
Furthermore, we propose a new advance sharing scheme for a stabilizer-based QSS, where a set of shares is an advance shareable set if and only if it is a forbidden set.
Therefore, our proposal is a scheme to maximize the advance shareable set in stabilizer-based QSS. 
Then, our proposal clarifies that it is possible to be advance shareable for a set of shares that is not be advance shareable by the previous scheme \cite{AdvanceSharing_Shibata}.

This paper is organized as follows.
In Section \ref{sec:2}, we review stabilizer codes and stabilizer-based QSS. 
In Section \ref{proof_of_forbidden}, we provide an example such that a set of shares is a forbidden set but not an advance shareable set by the previous scheme. 
In Section \ref{unitary_advance_sharing}, we propose an advance sharing scheme for stabilizer-based QSS, where 
a set of shares is advance shareable if and only if it is a forbidden set.
The conclusions follow in Section \ref{sec:four}.

\section{Preliminaries}\label{sec:2}
In this section, we review stabilizer codes and EAQECCs.
Throughout this paper, we suppose that $p$ is a prime number.
\subsection{Stabilizer codes}
Let $\{\ket{i}\mid i=0,\dots,p-1\}$ be an orthonormal basis for $p$-dimensional Hilbert space $\mathbb{C}^p$.
Let $\omega$ be a complex number such that is $\omega^p=1$ and $\omega^1,\omega^2,\dots,\omega^{p-1}$ are different.
We define two unitary matrices $X_p,Z_\omega$ 
that change $\ket{i}$ as 
$X_p \ket{i} = \ket{i+1\mod p}$ 
and $Z_\omega\ket{i} = \omega^i\ket{i}$ for $i=0,\dots,p-1$.
Consider the set $E_n = \{\omega^iX_p^{a_1}Z_\omega^{b_1}\otimes \dots\otimes X_p^{a_n}Z_\omega^{b_n}\mid i,a_j,b_j\in\{0,\dots,p-1\}\text{ for }j = 1,\dots,n\}$.
$E_n$ is a non-commutative finite group with matrix multiplication as its group operation. 
Denote by $\mathbb{F}_p$ the finite field with $p$ elements. 
For $\vec{a}=(a_1,\dots ,a_n)$ and $\vec{b} =(b_1,\dots ,b_n)\in\mathbb{F}_p^n$, we define 
$X_p(\vec{a}) = X_p^{a_1}\otimes\dots\otimes X_p^{a_n}$ 
and $Z_\omega(\vec{b})=Z_\omega^{b_1}\otimes\dots\otimes Z_\omega^{b_n}$.
We call a commutative subgroup of $E_n$ as a stabilizer. 

Suppose that eigenspaces of a stabilizer $S$ have dimension $p^k$. 
An $[[n,k]]_p$ quantum stabilizer code $Q(S)$ encoding $k$ qudits into $n$ qudits can be defined as 
a simultaneous eigenspace of all elements of $S$.

Now, we explain a way to describe a stabilizer $S$ by finite fields. 
For two vectors $(\vec{a}\mid \vec{b}),(\vec{c}\mid \vec{d})\in \mathbb{F}_p^{2n}$, the sympectic inner product is defined by
\begin{equation}
    \langle(\vec{a}\mid \vec{b}),(\vec{c}\mid \vec{d})\rangle_s=\langle\vec{a},\vec{d}\rangle_E-\langle\vec{b},\vec{c}\rangle_E, 
\end{equation}
where $\langle\cdot \mid \cdot\rangle_E$ is the Euclidean inner product.
For an $(n-k)$-dimensional $\mathbb{F}_p$-linear subspace $C$ of $\mathbb{F}_p^{2n}$, 
we define $C^{\perp} = \{\vec{a}\in\mathbb{F}^{2n}_p\mid \forall \vec{b}\in C,\langle\vec{a},\vec{b}\rangle_s=0\}$. 
We define $M(\vec{a}|\vec{b})$ as $M(\vec{a}|\vec{b}) = X_p(\vec{a})Z_\omega(\vec{b})\in {E_n}$ with $\vec{a}, \vec{b}\in \mathbb{F}_p^n$. 
We define a mapping $f(\omega^i M(\vec{a}|\vec{b}))$ from ${E_n}$ to $\mathbb{F}_p^{2n}$ by $f(\omega^i M(\vec{a}|\vec{b})) = (\vec{a}|\vec{b})$.
For a stabilizer $S$, $f(S)$ is an $\mathbb{F}_p$-linear space.

\subsection{Stabilizer-based QSS}
We review a stabilizer-based QSS \cite{QSS_Cleve_1999}. 
It is accomplished by the following steps:
\begin{algorithm}[H]
	\caption{Stabilizer-based QSS}
	\begin{algorithmic}[1]
    \STATE A dealer encodes a quantum secret by a stabilizer code.
    \STATE The dealer distributes each qudit of that codeword to a participant.
    \end{algorithmic}
\end{algorithm}

There are some procedures to reconstruct the secret for stabilizer-based QSS \cite{Unitary_reconstruction_Matsumoto_2017}.
One of the simplest procedures is to use erasure correction of the stabilizer code \cite{QSS_Cleve_1999}.
The access structure of a stabilizer-based QSS depends on the used stabilizer code. 

We review necessary and sufficient conditions for an index set $J\subset\{1,\dots,n\}$ to be a qualified set 
in QSS based on a stabilizer $S$ \cite{Unitary_reconstruction_Matsumoto_2017}. 
Shortening in this paper refers to making a new linear code $C'\subset \mathbb{F}^{2n-2}_p$ from a linear code $C\subset \mathbb{F}_p^{2n}$ 
by selecting vectors in $C$ where the $i$-th and the $(n+i)$th components $(1\leq i\leq n)$ are both zero and then eliminating the $i$-th and the $(n+i)$th components of the selected vectors.
Let $C_{(s)}^{(J)}$ be the code obtained by shortening the linear code $C$ for the element corresponding to the index set $J\subset\{1,\dots,n\}$.
Then, an index set $J$ is a qualified set if and only if the equation 
\begin{equation}
    f(S)_{(s)}^{(J)} = {f(S)^\perp}_{(s)}^{(J)}
\end{equation}
holds. 
In addition, an index set $J$ is a forbidden set if and only if its complement is a qualified set \cite{rampQSS_Ogawa_2005}.

EAQECC\cite{EAQECC_Brun_2006} is a class of quantum error-correcting codes.
We review a necessary and sufficient condition for an index set $J\subset\{1,\dots,n\}$ 
to be an advance shareable set 
in QSS based on a stabilizer $S$ by EAQECC\cite{AdvanceSharing_Shibata}. 
Then, the following lemma holds \cite{AdvanceSharing_Shibata}.
\begin{lemma}\label{junbi}
    Let $S$ be a stabilizer of ${E_{n}}$. 
    An index set $J\subset\{1,\dots,n\}$ is an advance shareable set 
    if and only if the equation
    \begin{equation}\label{4}
        \dim{f(S)_{(s)}^{(J)}}=\dim{f(S)}-2|J|
    \end{equation}
    holds.
\end{lemma}\hfill$\Box$

\subsection{Codewords of a stabilizer code}
We introduce the representation of codewords of a stabilizer code described in \cite{Unitary_reconstruction_Matsumoto_2017}.
Let $S$ be a stabilizer of $E_n$. 
Let $J$ be an index set $J\subset\{1,\dots,n\}$.
Let $\{\ket{\vec{v}_k}|\vec{v}_k\in\mathbb{F}_p^k\}$ be an  basis for $p^k$-dimensional Hilbert space.
We define $\ket{\Psi_{\vec{v}_k}}$ as the codeword of $Q(S)$ encoding $\ket{\vec{v}_k}$.
For a stabilizer $S$, we define $S_{(s)}^{(J)}$ as following:
\begin{equation}
    S_{(s)}^{(J)} = f^{-1}\left( f(S)_{(s)}^{(J)}\right).
\end{equation}
When $S$ is a stabilizer of $E_n$, $S_{(s)}^{(J)}$ becomes a stabilizer of $E_{n-|J|}$.
We define $\ell = \dim{Q\left(S_{(s)}^{(J)}\right)}$. 
We define an basis of $Q\left(S_{(s)}^{(J)}\right)$ as $\{\ket{\varphi_{\overline{J}}(\vec{u}_\ell)} \mid \vec{u}_\ell \in \mathbb{F}_p^\ell\}$.
Then, the following lemma holds \cite{Unitary_reconstruction_Matsumoto_2017}.
\begin{lemma}\label{Proving}
    Let $S$ be a stabilizer of $E_n$. 
    Let $J$ be an index set $J\subset\{1,\dots,n\}$.
    For ease of presentation, without loss of generality we may assume $\overline{J} = \{1,2\dots,|\overline{J}|\}$ and $J =\{|\overline{J}|+1,\dots,n\}$ by reordering indicies.
    If $J$ is a qualified set of QSS based on $S$, there exists an orthonormal basis $\left\{\ket{\phi_J\left(\vec{u}_\ell, \vec{v}_k\right)} \mid \vec{u}_\ell \in \mathbb{F}_p^\ell, \vec{v}_k \in \mathbb{F}_p^k\right\}$ of 
    $Q\left(S_{(s)}^{(\overline{J})}\right)$, and the following equation holds:
    \begin{equation}
        \ket{\Psi_{\vec{v}_k}} = \frac{1}{\sqrt{p^\ell}}\sum_{\vec{u}_\ell\in\mathbb{F}_p^\ell}\ket{\varphi_{\overline{J}}(\vec{u}_\ell)}\ket{\phi_J(\vec{u}_\ell,\vec{v}_k)}.
    \end{equation}
    \hfill$\Box$
\end{lemma}

\section{A relationship between forbidden sets and advance shareable sets for QSS constructed from EAQECC}\label{proof_of_forbidden}
In this section, we provide an example such that a set of shares is a forbidden set but not an advance shareable set for the previous scheme. 

In the advance sharing scheme of QSS based on a stablizer $S$ by EAQECC, there are cases where a set of shares is a forbidden set but not an advance shareable set.
We provide an example of such a case.

\begin{example}\label{ex_one}
    We define generators $\{M_1, M_2, M_3, M_4, M_5, M_6\}$ of a stabilizer $S$ of $E_7$ as follows:
    \begin{eqnarray}
        M_1 &=& X_2 \otimes X_2 \otimes X_2 \otimes X_2 \otimes I_2 \otimes I_2 \otimes I_2,\\
        M_2 &=& Z_{-1} \otimes Z_{-1} \otimes I_2 \otimes I_2 \otimes I_2 \otimes I_2 \otimes I_2,\\
        M_3 &=& I_2 \otimes I_2 \otimes Z_{-1} \otimes Z_{-1} \otimes I_2 \otimes I_2 \otimes I_2,\\
        M_4 &=& X_2 \otimes X_2 \otimes I_2 \otimes I_2 \otimes X_2 \otimes Z_{-1} \otimes Z_{-1},\\
        M_5 &=& I_2 \otimes I_2 \otimes X_2 \otimes X_2 \otimes Z_{-1} \otimes X_2 \otimes Z_{-1},\\
        M_6 &=& I_2 \otimes Z_{-1} \otimes Z_{-1} \otimes I_2 \otimes Z_{-1} \otimes X_2 \otimes X_2.
    \end{eqnarray}
    Then, an orthogonal basis of $f(S)$ is represented as follows:
    \begin{equation}
        \left\{
            \begin{array}{c}
                (1111000|0000000),\\
                (0000000|1100000),\\
                (0000000|0011000),\\
                (1100100|0000011),\\
                (0011010|0000101),\\
                (0000011|0110100)
            \end{array}
        \right\}.
    \end{equation}
    An basis of $f(S)^\perp$ is represented as follows:
    \begin{equation}
        \left\{
            \begin{array}{c}
                (1111000|0000000),\\
                (0000000|1100000),\\
                (0000000|0011000),\\
                (1100100|0000011),\\
                (0011010|0000101),\\
                (0000011|0110100),\\
                (0000100|0000010),\\
                (0000011|0000011)
            \end{array}
        \right\}.
    \end{equation}
    We define $J=\{5,6,7\}$, $\overline{J}=\{1,2,3,4\}$.
    Since 
    $f(S)_{(s)}^{(J)} = {f(S)^\perp}_{(s)}^{(J)}$ holds, $J$ is a qualified set and $\overline{J}$ is a forbidden set \cite{Unitary_reconstruction_Matsumoto_2017}.
    On the other hand, since the equation (\ref{4}) does not hold, 
    $\overline{J}=\{1,2,3,4\}$ is not an advance shareable set by the previous scheme \cite{AdvanceSharing_Shibata}. 
\end{example}

\section{Advance Sharing for Stabilizer-based QSS by unitary transformation}\label{unitary_advance_sharing}
We propose a new scheme of advance sharing for stabilizer-based QSS.
Let $S$ be a stabilizer of $E_n$.
Let an index set $J\subset\{1,\dots,n\}$ be a qualified set of QSS based on $S$.
For ease of presentation, without loss of generality we may assume $\overline{J} = \{1,2\dots,|\overline{J}|\}$ and $J =\{|\overline{J}|+1,\dots,n\}$ by reordering indicies.
From Lemma \ref{Proving}, 
there exist  bases $\left\{\ket{\phi_J\left(\vec{u}_\ell, \vec{v}_k\right)} \mid \vec{u}_\ell \in \mathbb{F}_p^\ell, \vec{v}_k \in \mathbb{F}_p^k\right\}$ of 
$Q\left(S_{(s)}^{(\overline{J})}\right)$, and the following equation holds.    
\begin{equation}
    \ket{\Psi_{\vec{v}_k}} = \frac{1}{\sqrt{p^\ell}}\sum_{\vec{u}_\ell\in\mathbb{F}_p^\ell}\ket{\varphi_{\overline{J}}(\vec{u}_\ell)}\ket{\phi_J(\vec{u}_\ell,\vec{v}_k)}.
\end{equation}

Let $\left\{ \ket{\vec{u}_\ell}|\vec{u}_\ell\in \mathbb{F}_p^\ell\right\}$ be an  basis for 
$p^\ell$-dimensional Hilbert space.
Since $\left\{\ket{\phi_J(\vec{u}_\ell,\vec{v}_k)}|\vec{u}_\ell\in\mathbb{F}_p^\ell,\vec{v}_k\in\mathbb{F}_p^k\right\}$ and 
$\left\{ \ket{\vec{u}_\ell}\ket{0}^{\otimes |J|-k-\ell}\ket{\vec{v}_k}|\vec{u}_\ell\in \mathbb{F}_p^\ell,\vec{v}_k\in \mathbb{F}_p^k\right\}$
are  bases with the same number of quantum states in them, we can define a unitary matrix $U_J$ sending $\ket{\vec{u}_\ell}\ket{0}^{\otimes |J|-k-\ell}\ket{\vec{v}_k}$ to $\ket{\phi_J(\vec{u}_\ell,\vec{v}_k)}$.

Here, we define a quantum secret $\ket{\psi_k}$ of $k$-qudits with complex coefficients $\alpha(\vec{v}_k)$ as follows:
\begin{equation}
    \ket{\psi_k} = \sum_{\vec{v}_k\in\mathbb{F}_p^k}\alpha(\vec{v}_k)\ket{\vec{v}_k}.
\end{equation}
Let \(\ket{\Psi(\psi_k)}\) denote the codeword of \(Q(S)\) encoding \(\ket{\psi_k}\), which can be expressed as follows:
\begin{equation}
    \ket{\Psi(\psi_k)} = \sum_{\vec{v}_k\in\mathbb{F}_p^k}\frac{\alpha(\vec{v}_k)}{\sqrt{p^\ell}}\sum_{\vec{u}_\ell\in\mathbb{F}_p^\ell}\ket{\varphi_{\overline{J}}(\vec{u}_\ell)}\ket{\phi_J(\vec{u}_\ell,\vec{v}_k)}.
\end{equation}
Therefore, the following equation holds:
\begin{align}
    \ket{\Psi(\psi_k)} =
    \left( I_{\overline{J}}\otimes U_J\right)\frac{1}{\sqrt{p^\ell}}\sum_{\vec{u}_\ell\in\mathbb{F}_p^\ell}\ket{\varphi_{\overline{J}}(\vec{u}_\ell)}\ket{\vec{u}_\ell}\ket{0}^{\otimes |J|-k-\ell}\ket{\psi_k}.
\end{align}
We define the initial state $\ket{\Phi_J}$ for a stabilizer $S$ and an index set $J$ as following:
\begin{equation}\label{initialState}
    \ket{\Phi_J} = \frac{1}{\sqrt{p^\ell}}\sum_{\vec{u}_\ell\in\mathbb{F}_p^\ell}\ket{\varphi_{\overline{J}}(\vec{u}_\ell)}\ket{\vec{u}_\ell}\ket{0}^{\otimes |J|-k-\ell}
\end{equation}
Then, the following equation holds:
\begin{equation}\label{13}
    \ket{\Psi(\psi_k)} = \left( I_{\overline{J}}\otimes U_J\right)\ket{\Phi_J}\ket{\psi_k}.
\end{equation}
Equation (\ref{13}) means that 
\(I_{\overline{J}}\) is the identity matrix applying on the qudits corresponding to \(\overline{J}\), the shares corresponding to \(\overline{J}\) are advance shareable.
Here, our proposal is accomplished by the following steps:
\begin{algorithm}[H]
	\caption{Advance Sharing for Stabilizer-based QSS by unitary transformation}
	\begin{algorithmic}[1]
    \STATE A dealer prepares the initial qudits $\ket{\Phi_J}$ in the equation (\ref{initialState}) for a stabilizer $S$ and an index set $J$.
    \STATE The dealer distributes the $j$-th qudit for all $j\in\overline{J}$ to participants in $\overline{J}$.
    \STATE The dealer applies $U_J$ on a $k$-qudit quantum secret $\ket{\phi_k}$ with the remaining qudits of \( \ket{\Phi_J} \).
    \STATE The dealer distributes each qudit obtained in Step 3 to the remaining participant.
    \end{algorithmic}
\end{algorithm}
    In this scheme, the shares of the complement of an advance shareable set are generated by applying unitary matrix $U_J$. 
    Therefore, the participants corresponding to the complement of an advance shareable set can reconstruct the secret by applying $U_J^{\dagger}$.
    Hence, the complement of an advance shareable set is always a qualified set. 
    Since the complement of a qualified set is a forbidden set \cite{rampQSS_Ogawa_2005}, 
    any advance shareable set is a forbidden set. 
    From Lemma \ref{Proving}, if \(\overline{J}\) is a forbidden set, then we can define \(U_J\). 
    Therefore, in our proposal, a set of shares is an advance shareable set if and only if it is a forbidden set.
\begin{remark}
    Suppose a sender wants to transmit a $k$-qudit quantum state $\ket{\phi_k}$ to a receiver.
    If the receiver pre-holds qudits corresponding to \( \overline{J} \) of the initial state $\ket{\Phi_J}$, and the sender transmits qudits corresponding to \( J \) of $\ket{\Psi(\psi_k)}$ as the codeword, this code becomes a $[[n-|\overline{J}|,k;|\overline{J}|]]+[[|\overline{J}|,\ell]]$ EAQECC \cite{EAQECC_from_Stabilizer_Brun_2012}.
\end{remark}
Here, we clarify an example where the proposed advance sharing scheme for a set of shares that is not advance shareable by the previous approach \cite{AdvanceSharing_Shibata}.
\begin{example}\label{ex_stabilizer}
    Let $S$ be the stabilizer defined in Example \ref{ex_one}.
    As shown in Example \ref{ex_one}, for \(J=\{5,6,7\}\) and \(\overline{J}=\{1,2,3,4\}\), 
    \(J\) is a qualified set and \(\overline{J}\) is a forbidden set.
    Let $\{\ket{\Psi_{0}},\ket{\Psi_{1}}\}$ be a orthonormal basis for $Q(S)$ as following:
    \begin{eqnarray*}
        \begin{array}{l}
            4\sqrt{2}\ket{\Psi_{0}} \\
            =\ket{0000000}+\ket{0000001}+\ket{0000010}+\ket{0000011}\\
            -\ket{0000100}-\ket{0000101}+\ket{0000110}+\ket{0000111}\\
            +\ket{1111000}+\ket{1111001}+\ket{1111010}+\ket{1111011}\\
            -\ket{1111100}-\ket{1111101}+\ket{1111110}+\ket{1111111}\\
            -\ket{0011000}+\ket{0011001}-\ket{0011010}+\ket{0011011}\\
            +\ket{0011100}-\ket{0011101}-\ket{0011110}+\ket{0011111}\\
            -\ket{1100000}+\ket{1100001}-\ket{1100010}+\ket{1100011}\\
            +\ket{1100100}-\ket{1100101}-\ket{1100110}+\ket{1100111},\\
            4\sqrt{2}\ket{\Psi_{1}} \\
            =\ket{0000000}-\ket{0000001}-\ket{0000010}+\ket{0000011}\\
            +\ket{0000100}-\ket{0000101}+\ket{0000110}-\ket{0000111}\\
            +\ket{1111000}-\ket{1111001}-\ket{1111010}+\ket{1111011}\\
            +\ket{1111100}-\ket{1111101}+\ket{1111110}-\ket{1111111}\\
            +\ket{0011000}+\ket{0011001}-\ket{0011010}-\ket{0011011}\\
            +\ket{0011100}+\ket{0011101}+\ket{0011110}+\ket{0011111}\\
            +\ket{1100000}+\ket{1100001}-\ket{1100010}-\ket{1100011}\\
            +\ket{1100100}+\ket{1100101}+\ket{1100110}+\ket{1100111}.\\
        \end{array}
    \end{eqnarray*}
    Then, the orthonormal basis of $Q\left(S_{(s)}^{(J)}\right)$, denoted by $\{\ket{\varphi_{\overline{J}}(\vec{u}_1)} \mid \vec{u}_1 \in \mathbb{F}_p\}$, is given as follows:
    \begin{eqnarray}
        \ket{\varphi_{\overline{J}}(0)} &=& \frac{1}{\sqrt{2}}\left(\ket{0000}+\ket{1111}\right)\\
        \ket{\varphi_{\overline{J}}(1)} &=& \frac{1}{\sqrt{2}}\left(\ket{0011}+\ket{1100}\right).\\
    \end{eqnarray}
    In addition, the orthonormal basis of $Q\left(S_{(s)}^{(\overline{J})}\right)$, denoted by $\left\{\ket{\phi_J\left(\vec{u}_1, \vec{v}_1\right)} \mid \vec{u}_1 \in \mathbb{F}_p, \vec{v}_1 \in \mathbb{F}_p\right\}$, is given as follows:
    \begin{eqnarray}
        2\sqrt{2}\ket{\phi_J\left(0, 0\right)} &=& \ket{000} + \ket{001} + \ket{010} + \ket{011} \nonumber\\
        &-& \ket{100} - \ket{101} + \ket{110} + \ket{111}\\
        2\sqrt{2}\ket{\phi_J\left(0, 1\right)} &=& \ket{000} - \ket{001} - \ket{010} + \ket{011} \nonumber\\
        &+& \ket{100} - \ket{101} + \ket{110} - \ket{111}\\
        2\sqrt{2}\ket{\phi_J\left(1, 0\right)} &=&-\ket{000} + \ket{001} - \ket{010} + \ket{011} \nonumber\\
        &+& \ket{100} - \ket{101} - \ket{110} + \ket{111}\\
        2\sqrt{2}\ket{\phi_J\left(1, 1\right)} &=&\ket{000} + \ket{001} - \ket{010} - \ket{011} \nonumber\\
        &+& \ket{100} + \ket{101} + \ket{110} + \ket{111}.
    \end{eqnarray}
    Then, the following equations hold.
    \begin{eqnarray}
        \ket{\Psi_{0}} &=& \frac{1}{\sqrt{2}}\left(\ket{\varphi_{\overline{J}}(0)}\ket{\phi_J\left(0, 0\right)} + \ket{\varphi_{\overline{J}}(1)}\ket{\phi_J\left(1, 0\right)}\right)\nonumber\\
        \ket{\Psi_{1}} &=& \frac{1}{\sqrt{2}}\left(\ket{\varphi_{\overline{J}}(0)}\ket{\phi_J\left(0, 1\right)} + \ket{\varphi_{\overline{J}}(1)}\ket{\phi_J\left(1, 1\right)}\right).\nonumber
    \end{eqnarray}
    Therefore, we define $U_J$ sending $\ket{\vec{u}_1}\ket{0}\ket{\vec{v}_1}$ to $\ket{\phi_J(\vec{u}_1,\vec{v}_1)}$ as follows:
\begin{align}
    \begin{autobreak}
        2\sqrt{2}U_J 
        = \big(\ket{000} + \ket{001} + \ket{010} + \ket{011} 
        - \ket{100} - \ket{101} + \ket{110} + \ket{111}  \big)\bra{000} 
        - \big(\ket{000} - \ket{001} + \ket{010} - \ket{011} 
        - \ket{100} + \ket{101} + \ket{110} - \ket{111}  \big)\bra{100} 
        + \big(\ket{000} - \ket{001} - \ket{010} + \ket{011} 
        + \ket{100} - \ket{101} + \ket{110} - \ket{111}  \big)\bra{001}
        + \big(\ket{000} + \ket{001} - \ket{010} - \ket{011} 
        + \ket{100} + \ket{101} + \ket{110} + \ket{111}  \big)\bra{101} 
        - \big(\ket{000} + \ket{001} - \ket{010} - \ket{011} 
        - \ket{100} - \ket{101} - \ket{110} - \ket{111}  \big)\bra{010} 
        + \big(\ket{000} + \ket{001} + \ket{010} + \ket{011} 
        + \ket{100} + \ket{101} - \ket{110} - \ket{111}  \big)\bra{011} 
        + \big(\ket{000} - \ket{001} + \ket{010} - \ket{011} 
        + \ket{100} - \ket{101} - \ket{110} + \ket{111}  \big)\bra{110} 
        + \big(\ket{000} - \ket{001} - \ket{010} + \ket{011} 
        - \ket{100} + \ket{101} - \ket{110} + \ket{111}  \big)\bra{111} .
    \end{autobreak}
\end{align}
Here, the following equations hold.
\begin{eqnarray*}
    \begin{array}{l}
        \ket{\Psi_{0}} \\
        =\frac{1}{\sqrt{2}}\left(I^{\otimes |\overline{J}|}\otimes U_J\right)\Big(\ket{\varphi_{\overline{J}}(0)}\ket{000}+\ket{\varphi_{\overline{J}}(1)}\ket{010}\Big)\\
        \ket{\Psi_{1}} \\
        =\frac{1}{\sqrt{2}}\left(I^{\otimes |\overline{J}|}\otimes U_J\right)\Big(\ket{\varphi_{\overline{J}}(0)}\ket{001}+\ket{\varphi_{\overline{J}}(1)}\ket{011}\Big).
    \end{array}
\end{eqnarray*}
Then, the initial state $\ket{\Phi_J}$ is following:
\begin{equation}
    \ket{\Phi_J} = \frac{1}{\sqrt{2}}\big(\ket{\varphi_{\overline{J}}(0)}\ket{00}+\ket{\varphi_{\overline{J}}(1)}\ket{01}\big)
\end{equation}
Therefore, for any 1-qubit state \(\ket{\psi_1}\), the codeword of \(Q(S)\) encoding \(\ket{\psi_1}\) can be expressed as the following equation:
\begin{eqnarray*}
    \begin{array}{l}
    \ket{\Psi(\psi_1)} =\left(I^{\otimes |\overline{J}|}\otimes U_J\right)(\ket{\Phi_J}\ket{\psi_1}).
    \end{array}
\end{eqnarray*}
This equation means that we can distribute the first 4 qubits of the initial state $\ket{\Phi_J}$ before the secret \(\ket{\psi_1}\) is determined.
\end{example}

\section{Conclusion}\label{sec:four}
In this paper,  we propose a new advance sharing scheme for stabilizer-based QSS, 
where a set of shares is an advance shareable set if and only if it is a forbidden set.
It is known that a set of shares is an advance shareable set only if it is a forbidden set. 
Therefore, our proposal is a scheme to maximize the advance shareable set in stabilizer-based QSS. 
Then, our proposal clarifies that it is possible to be advance shareable for a set of shares that is not be advance shareable by the previous scheme.

\section*{Acknowledgments}
The author would like to thank Professor Ryutaroh Matsumoto for helpful advice. 

\bibliographystyle{plain}
\bibliography{papers_liblary}

\end{document}